\documentstyle[aps,prd,eqsecnum,twocolumn,epsf]{revtex}

\newcommand{\beq}{\begin{equation}}
\newcommand{\beqn}{\begin{eqnarray}} 
\newcommand{\eeq}{\end{equation}}
\newcommand{\eeqn}{\end{eqnarray}}
\newcommand{\beqa}{\begin{eqnarray}}
\newcommand{\eeqa}{\end{eqnarray}}

\newcommand{\lsim}{\mbox{\raisebox{-1.ex}{$\stackrel
     {\textstyle<}{\textstyle \sim}$}}}
\newcommand{\square}{\kern1pt\vbox{\hrule height
1.2pt\hbox{\vrule width 1.2pt\hskip 3pt
   \vbox{\vskip 6pt}\hskip 3pt\vrule width 0.6pt}\hrule
height 0.6pt}\kern1pt}

\def\beq{\begin{equation}}

\def\eeq{\end{equation}}

\begin{document}

\draft
\twocolumn[\hsize\textwidth\columnwidth\hsize\csname
@twocolumnfalse\endcsname

\title{{\bf Density perturbations in the Ekpyrotic Universe and 
string-inspired generalizations}} \author{Shinji Tsujikawa} 
\address{Research Center for the Early Universe, University of Tokyo, 
Hongo, Bunkyo-ku, Tokyo 113-0033, Japan \\[.3em]} \date{\today} \maketitle
\begin{abstract}
We study density perturbations in several cosmological models motivated by 
string theory.  The evolution and the spectra of curvature perturbations 
${\cal R}$ are analyzed in the Ekpyrotic scenario
and nonsingular string cosmologies.  We find that these string-inspired models
generally exhibit blue spectra in contrast to standard slow-roll 
inflationary scenarios. We also clarify the parameter range where ${\cal R}$ 
is enhanced on superhorizon scales.
\end{abstract}
\vskip 2pc
]

\section{Introduction}                           %

The pre-big-bang (PBB) cosmological models based on the low-energy effective 
actions of string theory have been widely studied as possible candidates to 
describe the evolution of the Universe at very high density \cite{review}.  
In the PBB scenario there exist two disconnected branches, one of which 
corresponds to the stage of superinflation driven by the kinetic term of 
the dilaton field and another of which is the Friedmann branch of 
decelerating expansion \cite{PBB,GVdilaton}.  While it is difficult to 
avoid the big bang singularity in the tree-level string action 
\cite{BV,tree}, two branches can be smoothly joined to
each other by taking into account quantum loop corrections.

Among the nonsingular string-inspired scenarios, the model proposed by 
Antoniadis et al.  \cite{oneloop} involves the Gauss-Bonnet curvature 
invariant coupled to the modulus field, which enables the big bang 
singularity to be avoided even in the closed \cite{closed} and the anisotropic 
Bianchi type-I Universes \cite{yaji}.  In the presence of the higher-order 
derivatives and the Gauss-Bonnet term coupled to the dilaton field, it was 
found in refs.~\cite{GMV,BM} that the pre-big-bang inflationary evolution 
is followed by a phase with decreasing curvature without a singularity (see 
also \cite{FMS,CCM,BEM}).  Recently a new scenario --the Ekpyrotic 
model -- has been proposed by Khoury et al. \cite{KOST,KOST2}.  
According to this scenario, the collision of two parallel branes imbedded in the 
extra-dimensional bulk signals the beginning of the hot, expanding,
big bang of the standard cosmology.
Prior to the collision the Universe was slowly contracting.

It should be possible to discriminate the viability of these models by 
evaluating the spectra of density perturbations. 
 In the PBB scenario without loop 
corrections, the curvature perturbation ${\cal R}$ is blue-tilted with spectral 
index $n \simeq 4$ \cite{Bru}, which contradicts with the observationally 
supported flat spectra with $n \simeq 1$.  
In order to make contact with observations precisely, 
it is very important to extend to nonsingular cosmological models with
a successful graceful exit.
 In what follows we will analyze 
the spectra of curvature perturbations in the nonsingular models and the 
Ekpyrotic scenario.  We will also clarify the cases where ${\cal R}$ 
is amplified on superhorizon scales even in the one-field 
model.  This should again be contrasted with standard single-field 
inflationary models where ${\cal R}$ is constant on large scales.

 \section{Perturbed equations}

Let us consider the following four-dimensional action 
\begin{eqnarray}
 S = \int d^4 x \sqrt{-g} \Bigl[ \frac12 f(R, \phi) &-& \frac12 \omega 
 (\phi) (\nabla \phi)^2 \nonumber \\
&-& V(\phi)
 +{\cal L}_c \Bigr],
\label{lag}
\end{eqnarray}
where $f(R, \phi)$ is a function of the Ricci scalar $R$ and
a scalar field $\phi$.  $\omega(\phi)$ and $V(\phi)$ are general functions
of $\phi$.  The Lagrangian ${\cal L}_c$ represents the higher-order
corrections to the tree-level action.  The action (\ref{lag}) includes a wide 
variety of theories, e.g., Einstein gravity, scalar tensor theories, and 
the low-energy effective string theories reduced from the 
higher-dimensional actions.  In what follows we shall focus on the models 
inspired from string theories.

While it is difficult to avoid the big bang singularity 
in the PBB scenario using only the tree-level action, 
the existence of loop corrections is capable to overcome such singularity 
problem.  In this work we shall consider the higher-order corrections, 
given by 
\begin{eqnarray}
 {\cal L}_c = -\frac12 \alpha' \lambda
  \xi(\phi) \left[ c R_{\rm GB}^2+ d 
 (\nabla \phi)^4 \right],
\label{lagc}
\end{eqnarray}
where $R_{\rm GB}^2 =R^2-4R^{\mu\nu}R_{\mu\nu}+ 
R^{\mu\nu\alpha\beta}R_{\mu\nu\alpha\beta}$ is the 
Gauss-Bonnet term.  Hereafter the inverse string 
tension, $\alpha'$, is set to unity.

A perturbed space-time
metric has the following form for scalar perturbations 
in an arbitrary gauge:
\begin{eqnarray}
ds^2 = &-& (1+2A)dt^2 + 2a(t)B_{,i} dx^idt \nonumber \\
&+& a^2(t)[(1-2\psi)\delta_{ij}+2E_{,i,j}] dx^i dx^j, 
\label{pmetric}
\end{eqnarray}
where $a(t)$ is the scale factor, and a comma means usual flat space 
coordinate derivative.  It is convenient to introduce gauge-invariant 
variable, defined by 
\begin{eqnarray}
{\cal R} \equiv \psi+\frac{H}{\dot{\phi}}\delta\phi,
\label{metric}
\end{eqnarray}
where $H \equiv \dot{a}/a$ is the Hubble expansion rate. 
This quantity corresponds to the curvature perturbation in the uniform curvature gauge.  
The perturbed Einstein equations for scalar perturbations are 
written in the form \cite{Hwang_gau,CHC} 
\begin{eqnarray}
\frac{1}{a^3Q} \left(a^3Q {\cal R} \right)^{\bullet}
-s \frac{\Delta}{a^2} {\cal R}=0,
\label{peinstein}
\end{eqnarray}
where
\begin{eqnarray}
Q \equiv \frac{\omega \dot{\phi}^2 +3I
(\dot{F}-4\lambda c \dot{\xi}H^2)-6\lambda d \xi 
\dot{\phi}^4}{\left( H + I\right)^2},
\label{Q}
\end{eqnarray}
\begin{eqnarray}
s \equiv 1+\frac{4\lambda c\xi \dot{\phi}^4
-16\lambda c \dot{\xi}\dot{H}I+ 8\lambda c (\ddot{\xi}-\dot{\xi}
H)I^2} {\omega \dot{\phi}^2 +3I(\dot{F} -4\lambda c 
\dot{\xi}H^2)-6\lambda d \xi \dot{\phi}^4}.
\label{s}
\end{eqnarray}
Here $F$ and $I$ are defined as $F=\partial f/\partial R$ and $I \equiv 
(\dot{F}-4\lambda c \dot{\xi}H^2)/ (2F-8\lambda c \dot{\xi} H)$.

Let us introduce a new quantity, $\Psi \equiv z{\cal R}$, with
$z \equiv a\sqrt{Q}$.  Then the each Fourier component of $\Psi$
satisfies the second order differential equation
\begin{eqnarray}
\Psi_k''+\left(sk^2-\frac{z''}{z}\right)\Psi_k=0,
\label{Psi}
\end{eqnarray}
where a prime denotes the derivative with respect to conformal time,
$\eta=\int a^{-1}dt$.
In the large scale limit, $|sk^2| \ll |z''/z|$, eq.~(\ref{Psi}) is easily 
integrated to give
\begin{eqnarray}
{\cal R}_k =C_k+D_k \int \frac{d\eta}{z^2},
\label{Rk}
\end{eqnarray}
where $C_k$ and $D_k$ are integration constants
corresponding to the growing and decaying mode, respectively.  
The curvature perturbation is conserved on superhorizon 
scales as long as the decaying mode is not strongly dominating,
as in the case of the single field, slow-roll inflationary scenarios.  A 
counter example was recently found in ref.~\cite{LS} in the context of 
fast-roll inflation.  

In the string-inspired models, when the evolution of $z$ 
before the bounce is given in the form 
\begin{eqnarray}
z \propto (-\eta)^{\alpha},
\label{z}
\end{eqnarray}
the second term in eq.~(\ref{Rk})
yields $\int d\eta/z^2 \propto (-\eta)^{1-2\alpha}$.
Therefore ${\cal R}_k$ can be amplified for $\alpha \ge 1/2$ on superhorizon 
scales, while it is not for $\alpha<1/2$ (Note that ${\cal R}_k \propto {\rm ln}
 (-\eta)$ for $\alpha=1/2$).
  Whether this enhancement occurs or not depends on the models of 
 string theory as we will show later.

In order to obtain the spectra of curvature perturbations, it is required 
to go to the next order solution of eq.~(\ref{Psi}). 
If $s$ is a positive constant, the solution for $\Psi_k$ is expressed by the 
combination of the Hankel functions: 
\begin{eqnarray}
\Psi_k=\frac{\sqrt{-\pi \eta}}{2}\left[ c_1 H_{\nu}^{(1)}
(x) +c_2 H_{\nu}^{(2)}(x) \right],
\label{han}
\end{eqnarray}
where $x \equiv -\sqrt{s} k\eta$ and $\nu \equiv |1-2\alpha|/2$.  
Making use of the relation $H_{\nu}^{(2,1)} (-\sqrt{s}k\eta) \to \pm (i/\pi) 
\Gamma(\nu) (-\sqrt{s}k\eta/2)^{-\nu}$ for long wavelength perturbations 
($-\sqrt{s}k\eta \to 0$), we get the spectrum of curvature perturbations as 
\begin{eqnarray}
P_{{\cal R}} \equiv \frac{k^3}{2\pi^2}\left| 
{\cal R}_k \right|^2 \propto k^{3-2\nu},
\label{PR}
\end{eqnarray}
in which case the spectral tilt is 
\begin{eqnarray}
n-1=3-\left| 1-2\alpha \right|.
\label{ind}
\end{eqnarray}

If the corrections ${\cal L}_c$ are not taken into account, 
$s$ is exactly unity as in the case of the Einstein gravity.  
In the context of 
nonsingular string cosmologies, $s$ is generally the time-varying function 
and even can change sign due to the presence of the higher-order 
corrections.  In such cases the formula (\ref{ind}) can not be directly 
applied, but it is still valid if $s$ is slowly varying positive function.
In the next section we shall analyze the evolution and the spectra of curvature 
perturbations in several different models.

 \section{String inspired models}

%
\subsection{Ekpyrotic Universe}

The Ekpyrotic scenario is based on the collision of two parallel branes 
imbedded in the extra-dimensional bulk.  The position of the moving brane 
before the collision is represented by a scalar field $\phi$, 
with negative exponential potential 
\begin{eqnarray}
V(\phi)=-V_0 e^{-\sqrt{2/p} \phi}.
\label{ekp_po}
\end{eqnarray}
In the original version of the Ekpyrotic 
scenario \cite{KOST}, the collision of branes takes place at some finite 
value of $\phi$. 
The effective action in this model corresponds to $f=R$, $\omega=1$, 
$V \ne 0$, and ${\cal L}_c=0$, in which case one has $Q=\dot{\phi}^2/H^2$ and 
$s=1$.  The background evolution before the bounce ($t<0$) is given by 
\begin{eqnarray}
a \propto (-t)^p, ~~H=\frac{p}{t},~~
\dot{\phi}^2=\frac{2p}{t^2},~~V=\frac{p(3p-1)}{t^2}.
\label{back}
\end{eqnarray}
Therefore we find  $z=a\dot{\phi}/H \propto (-t)^p 
\propto (-\eta)^{p/(1-p)}$, which means $\alpha=p/(1-p)$.
Since a very slow contraction leads to the condition, 
$0<p \ll 1/3$, $\alpha$ is constrained to be $0<\alpha \ll 1/2$.
Then the curvature perturbation is conserved on large scales due to the 
negligible contribution of the decaying mode of eq.~(\ref{Rk}).

Since $s$ is exactly unity in this case, the result (\ref{ind}) 
can be applied to give  
\begin{eqnarray}
n-1=\frac{2}{1-p}.
\label{ekiind}
\end{eqnarray}
Therefore one has the blue spectrum $n \simeq 3$ for $p \simeq 0$ ($\alpha 
\simeq 0$).  This means the difficulty to generate the scale-invariant 
spectrum in the old Ekpyrotic scenario, which supports the results in 
refs.~\cite{Lyth,BF,Hwang_ekp}.  

In the new Ekpyrotic scenario \cite{KOST2}, it is assumed that the brane 
collision occurs at the field value $\phi=-\infty$ with vanishing potential 
$V \simeq 0$.  This corresponds to the case with $p=1/3$ and $\alpha=1/2$,
which indicates that ${\cal R}_k$ is logarithmically divergent at the bounce 
($\eta=0$).  On the other hand the gravitational potential $\Phi_k$ 
is finite and even exhibits nearly flat spectra with small 
amplitudes \cite{KOST2}.  In the case where ${\cal R}$ is divergent and 
$\Phi$ is much smaller than unity, however, we cannot choose any 
appropriate time-displacement $\delta t$ in gauge transformations 
\cite{Lyth2}.  Therefore the cosmological perturbation theory is no longer valid 
at the bounce, which suggests the difficulty to produce the observationally 
supported curvature perturbations.  This situation is similar to the 
tree-level PBB scenario where ${\cal R}_k$ grows logarithmically toward the 
singularity at $\eta=0$ and the spectrum of the curvature perturbation is 
far from scale-invariant.

\subsection{Modulus-driven inflation}

Let us next consider a model of $f=R$, $\omega=1$, $V=0$,
$c=-1$, $d=0$, and $\xi={\rm ln}[2e^{\phi}\eta^4(ie^{\phi})]$ with 
$\eta(ie^{\phi})$ being the Dedekind $\eta$ 
function \cite{oneloop}. The singularity problem which plagues the PBB 
scenario can be avoided by taking into account the above form of the 
one-loop correction in the Einstein frame with modulus ($\phi$) and dilaton 
fields.  Since the evolution of the solution is mainly determined by the 
motion of the modulus field, we shall focus on the one-field system with 
positive $\lambda$ whereby singularity can be avoided.
 
Starting from a large initial value, $|\phi| \gg 1$,
the Universe exhibits superinflation ($\dot{H}>0$) until the graceful exit 
at $t=0$.  During this stage, the background evolution is 
given by \cite{yaji,KS} 
\begin{eqnarray}
a \simeq a_0,~~~H =\frac{H_0}{t^2},~~~
\dot{\phi}=\frac{5}{(-t)},
\label{backevo}
\end{eqnarray}
 where $a_0$ and $H_0$ are constants. 
 After the Hubble parameter reaches its peak value 
 $H=H_{\rm max}$, the system connects to the Friedmann-like
 Universe with $H\simeq 1/(3t)$.
 
 During the modulus-driven inflation, one has $\dot{\xi}\simeq
 -(2\pi/3)\dot{\phi}\sin {\rm h}\phi\simeq (\pi/3)\dot{\phi}
 e^{-\phi} \propto (-t)^4$.  
 Therefore the value $Q$ defined by (\ref{Q}) 
 evolves as $Q \propto (-t)^2$, which leads to 
 $z \propto (-t) \propto (-\eta)$.  
 Since $\alpha=1$ in this case, the curvature perturbation is enhanced 
 on superhorizon scales during superinflation 
 due to the growth of the second term in eq.~(\ref{Rk}).  Even in the 
 one-field model the conservation of ${\cal R}_k$ can be violated after 
 horizon crossing when $\alpha$ is larger than $1/2$.  
 We should keep in mind that the conservation of ${\cal R}_k$ is
 the specifics of the slow-roll inflationary scenarios 
 where the strongly dominated decaying mode is excluded
 (see refs.~\cite{LS,STY}).
 
 In the present model $s$ is much larger 
than unity during modulus-driven inflation and is proportional to
$(-t)$, in which case eq.~(\ref{ind}) is not applicable.  
{}From eq.~(\ref{peinstein}) we easily find that $R_k$ satisfies 
\begin{eqnarray}
\ddot{{\cal R}}_k+\frac{2}{t}\dot{{\cal R}}_k-\beta
\frac{k^2}{a_0^2} t {\cal 
R}_k=0,
\label{dR_k}
\end{eqnarray}
where $\beta~(>0)$ is a constant that depends on $H_0$.  The solution of this 
equation is expressed in terms of the Bessel functions with $x \equiv 
\frac23 \sqrt{\beta} \frac{k}{a_0}(-t)^{3/2}$: 
\begin{eqnarray}
{\cal R}_k=(-t)^{-1/2} \left[c_1 J_{-1/3}(x)+c_2J_{1/3}(x)\right],
\label{Bessel}
\end{eqnarray}
which asymptotically approaches the Minkowski vacuum
for $x \to \infty$.  Making use of the relation $J_{\pm 1/3}
(x) \propto k^{\pm 1/3}$ in the $x \to 0$ limit, 
the spectrum of the long wave curvature perturbation is 
\begin{eqnarray}
P_{{\cal R}} \propto k^{7/3}.
\label{PowerM}
\end{eqnarray}
This is a blue spectrum with spectral index $n=10/3$.
While  the curvature perturbation grows on superhorizon scales, 
its spectrum is not scale invariant.
In order to generate the scale invariant spectra, one requires 
the power exponent of $z$ close to $\alpha=2$ or $\alpha=-1$.

\begin{figure}
\epsfxsize = 3.5in
\epsffile{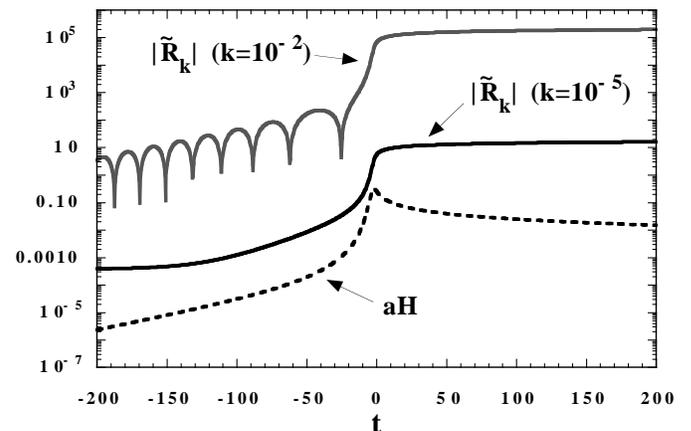} 
\caption{
The evolution of the quantity $aH$ and the curvature perturbation 
$|\tilde{\cal R}_k|=|k^{3/2}{\cal R}_k|$ for two different momenta during 
the modulus-driven inflation and the subsequent Friedmann-like Universe.  We 
choose $\lambda=6/\pi$, $\sigma=0$, and $\dot{\sigma}=0.2$ at $t=0$, whereby 
singularity is avoided.  }
\label{modulus}
\end{figure}

In Fig.~\ref{modulus} we plot the evolution of 
the curvature perturbation for two modes $k=10^{-5}$ and $10^{-2}$.  
The mode $k=10^{-5}$ crosses the horizon ($k=aH$) around $t=-150$, 
after which ${\cal R}_k$ continues to be enhanced until the graceful exit and freezes 
for $t>0$.  The mode $k=10^{-2}$ stays inside the horizon until 
$t~\lsim~-10$, during which the curvature perturbation exhibits some growth 
with oscillations ${\cal R}_k \propto (-t)^{-5/4}\cos(x)$ as found by 
eq.~(\ref{Bessel}).  After horizon crossing the evolution of ${\cal 
R}_k$ is similar to the $k=10^{-5}$ case.

It was claimed in ref.~\cite{PP}
that the bouncing Universe generally exhibits singular behavior of
the gravitational potential $\Phi_k$.
This result can not be directly applied to our model 
since the Universe is slowly expanding due to the presence
of higher-order quantum corrections. 
In fact the curvature perturbation ${\cal R}_k$ as well as
the background quantities are finite at the graceful exit, which 
means that the gravitational potential remains finite from 
eq.~(\ref{metric}). Therefore the cosmological perturbation theory
can be valid as long as the fluctuations do not exceed 
the linear level. This is contrasted with the new Ekpyrotic scenario
where ${\cal R}_k$ is divergent at the bounce.

\subsection{Dilaton-driven inflation}

Nonsingular cosmological solutions have been also found in 
refs.~\cite{GMV,BM,FMS} by including the higher order derivatives 
and loop corrections.  The simplest version corresponds to the case with 
$f=e^{-\phi}R$, $\omega=\xi=-e^{-\phi}$, $V=0$, $c=-1$, $d=1$, and 
$\lambda=-1/4$, which we shall analyze below. Notice that the PBB scenario 
does not include the $\alpha'$ correction in eq.~(\ref{lag}), in which case 
the background evolution is described by $a \propto 
(-\eta)^{(1-\sqrt{3})/2}$ and $e^{\phi}= (-\eta)^{-\sqrt{3}}$ in the string 
frame \cite{GVdilaton}.  Then one has $z \propto (-\eta)^{1/2}$, yielding 
the blue spectrum, $n=4$.

\begin{figure}
\epsfxsize = 3.5in
\epsffile{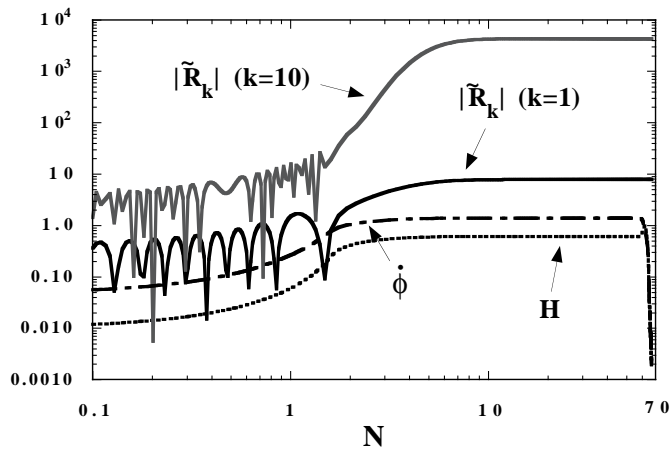} 
\caption{The evolution of the curvature perturbation 
$|\tilde{\cal R}_k|=|k^{3/2}{\cal R}_k|$ as a function of e-folds 
for two different momenta during the 
dilaton-driven inflation and the graceful exit.  
We choose the initial 
value of dilaton to be $\phi=-140$, in which case one has sufficient amount 
of inflation ($N>60$).  We also plot the evolution of $\dot{\phi}$ and $H$, 
which are nearly constant during the string phase.  Due to the presence of loop 
corrections, singularity avoidance is realized around $N=60$.}
\label{dilaton}
\end{figure}

In the presence of the $\alpha'$ correction, the initial low-curvature 
dilaton-driven phase is followed by the string phase with linearly 
growing dilaton and nearly constant Hubble parameter (see 
Fig.~\ref{dilaton}).  The fixed values during this stage are $\dot{\phi}_f 
\simeq 1.40$ and $H_f \simeq 0.62$, leading to the sufficient amount of 
inflation with e-folds $N \equiv {\rm ln} (a/a_i) >60$ provided that the 
dilaton field satisfies $|\phi| \gg 1$ initially \cite{CHC}.  In 
Fig.~\ref{dilaton} inclusion of the loop corrections makes it possible to 
connect smoothly to the Friedmann branch around $N \simeq 60$.

During the string phase with constant $\dot{\phi}$ and $H$, one easily 
finds that $a \propto (-\eta)^{-1}$ and $Q \propto e^{-\phi}
\propto (-\eta)^{\dot{\phi}_f/H_f}$, which yields 
$\alpha=-1+\dot{\phi}_f/(2H_f)$.  Therefore, when $s$ is 
positive constant, the spectral tilt of the large scale curvature 
perturbation is $n-1=3-|3-\dot{\phi}_f/H_f| \simeq 2.26$.  It was 
shown in ref.~\cite{CCM} that the ratio $\dot{\phi}_f/H_f$ is required to 
range in the region $2<\dot{\phi}_f/H_f<3$ for the successful graceful exit 
in the presence of other forms of $\alpha'$ corrections such as 
$G^{\mu\nu}\partial_{\mu}\phi\partial_{\nu}\phi$ and $\square \phi 
(\partial_{\mu}\phi)^2$.  While it is possible to have sufficient amount of 
inflation ($N>60$) in these scenarios, the spectral indices are 
constrained to be $3<n<4$ for the constant positive $s$, implying the 
difficulty to generate the flat spectra sourced by the dilaton fluctuation.

In the present model with fixed points $\dot{\phi}_f \simeq 1.40$ and $H_f 
\simeq 0.62$, $s$ changes sign during a short transition 
from the dilaton-driven to the string phase, after which $s$
is approximately negative constant until the graceful exit.
In this case the solution for eq.~(\ref{Psi}) can be written in the form 
\begin{eqnarray}
\Psi_k=\sqrt{-\eta} 
\left[c_1 I_{\nu} (x) +c_2 K_{\nu} (x) \right],
\label{Psid}
\end{eqnarray}
where $x=-\sqrt{|s|}k\eta$.  
Here $I_{\nu}$ and $K_{\nu}$ are modified 
Bessel functions, whose asymptotic solutions are $I_{\nu} \propto k^{\nu}$, 
$K_{\nu} \propto k^{-\nu}$ for $x \to 0$, and $I_{\nu} \sim e^x/\sqrt{2\pi 
x}$, $K_{\nu} \sim \sqrt{\pi/(2x)}e^{-x}$ for $x \to \infty$.  Therefore in 
the large scale limit ($|sk^2| \ll |z''/z|$), one reproduces the 
spectrum $P_{\cal R} \propto k^{\dot{\phi}_f/H_f}$ as in the case of the 
constant positive $s$.  For the modes which are inside the horizon, the 
curvature perturbation exhibit exponential increase, after which ${\cal R}_k$
is frozen as is found in Fig.~\ref{dilaton}.  
Since $\alpha=-1+\dot{\phi}_f/(2H_f)$ is smaller than $1/2$, the curvature 
perturbation is frozen after the horizon crossing.
This conservation is similar to the standard slow-roll inflationary scenarios, 
but the spectrum of the density perturbation is highly blue-tilted due to the 
enhancement of small scale fluctuations inside the horizon.
It is also worth mentioning that at the graceful exit  ${\cal R}_k$ 
remains finite together with the gravitational potential (see 
Fig.~\ref{dilaton}).

 \section{Summary and discussions}

\begin{figure}
\epsfxsize = 3.5in
\epsffile{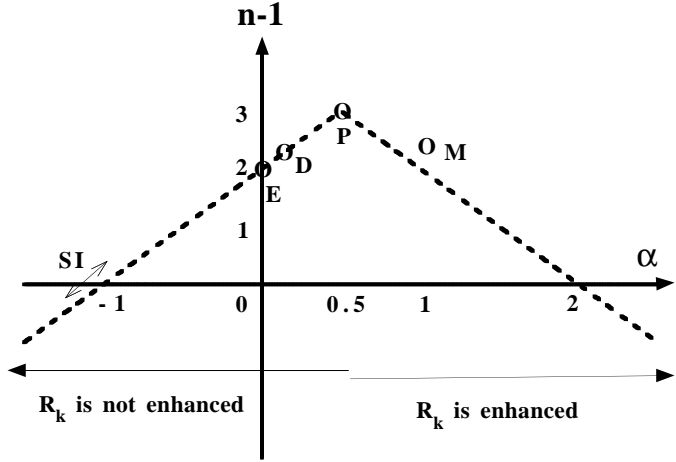} 
\caption{The spectral tilts in string-inspired cosmological models.  
``P, E, M, D'' correspond to the case of the pre-big-bang scenario, 
Ekpyrotic Universe, modulus-driven inflation with a Gauss-Bonnet term, 
and dilaton-driven inflation with higher-order corrections, respectively.  
``SI'' denotes the standard slow-roll inflation.  }
\label{spectrum}
\end{figure}

We shall summarize the density perturbation spectra 
in Fig.~\ref{spectrum}.
The quantity $z=a\sqrt{Q}$ is an important one to determine the spectra and 
the evolution of curvature perturbations.  When $z \propto (-\eta)^\alpha$ 
and the quantity $s$ is time-independent, the spectral tilts in the large 
scale limit ($|sk^2| \ll |z''/z|$) are given by eq.~(\ref{ind}), 
which are plotted by 
the dotted line in Fig.~\ref{spectrum}.  We also show the region where the 
curvature perturbation is enhanced on superhorizon scales 
($\alpha \ge 1/2$).

The simplest version of the pre-big bang scenario corresponds to the case of 
$\alpha=1/2$ and $n-1=3$.  In the Ekpyrotic Universe or the 
dilaton-driven inflation with higher-order corrections, the spectral 
tilts are $2< n-1 \le 3$ with $0< \alpha \le 1/2$.  In the 
modulus-driven inflation with a Gauss-Bonnet term, one has $\alpha=1$
and $n-1=7/3$.  In this case the spectral index is not on the line of 
$n-1=3-|1-2\alpha|$ due to the fact that $s$ is the time-varying function.  
In addition the curvature perturbation exhibits parametric amplification on 
superhorizon scales since $\alpha$ is larger than 1/2.

In string-inspired models considered in this work, we obtain blue spectra 
with $n>3$.  In order to generate the scale-invariant spectra ($n \simeq 1$), 
the background evolution is required to be $\alpha \simeq -1$ 
or $\alpha \simeq 2$.  The standard slow-roll inflation corresponds to the 
former case with conserved curvature perturbations after horizon crossing,
while the latter one is the case where ${\cal R}$ is enhanced on superhorizon 
scales.  It is certainly of interest to investigate whether there exist some 
models motivated by string theory which lead to the spectral indices 
with $|n-1|~\lsim~1$.

Even if the single-field string-inspired scenarios have 
some difficulty to generate the flat spectra, the quantum 
fluctuation of a light scalar field such as the axion 
may originate the large-scale curvature perturbations \cite{CEW}.  
In fact it was recently claimed in ref.~\cite{LW} that 
scale-invariant curvature perturbations may be obtained
after inflation if a scalar field 
called ``curvaton'' produces an almost flat spectrum of isocurvature 
perturbations due to a light mass during inflation
(see also refs.~\cite{ES,MT}).
The applications of this idea to the string-inspired models are left to 
future work.

\section*{ACKOWLEDGMENTS}
The author thanks Bruce Bassett, Robert Brandenberger, 
Cyril Cartier, Alexei Starobinsky, and Jun'ichi Yokoyama for useful 
discussions.  He is also thankful for financial support from the JSPS (No.  
04942).  


\end{document}